\pdfoutput=1
\documentclass[preprint,preprintnumbers,amsmath,amssymb,floatfix,endfloats*]{revtex4}

\usepackage{graphicx}
\usepackage{dcolumn}
\usepackage{bm}
\usepackage{amsfonts}

\DeclareMathAlphabet{\mathsfsl}{OT1}{cmr}{bx}{it}
\begin{document}
\title{Shear band healing in amorphous materials by small-amplitude oscillatory shear deformation}
\author{Nikolai V. Priezjev$^{1,2}$}
\affiliation{$^{1}$Department of Mechanical and Materials
Engineering, Wright State University, Dayton, OH 45435}
\affiliation{$^{2}$National Research University Higher School of
Economics, Moscow 101000, Russia}
\date{\today}
\begin{abstract}

The effect of small-amplitude periodic shear on annealing of a shear
band in binary glasses is investigated using molecular dynamics
simulations.  The shear band is first introduced in stable glasses
via large-amplitude periodic shear, and then amorphous samples are
subjected to repeated loading during thousands of cycles at strain
amplitudes below the yield strain.  It was found that with
increasing strain amplitude, the glasses are relocated to deeper
potential energy levels, while the energy change upon annealing is
not affected by the glass initial stability.  The results of
mechanical tests indicate that the shear modulus and yield stress
both increase towards plateau levels during the first few hundred
cycles, and their magnitudes are largest when samples are loaded at
strain amplitudes close to the yield strain. The analysis of
nonaffine displacements reveals that the shear band breaks up into
isolated clusters that gradually decay over time, leading to nearly
reversible deformation within the elastic range. These results might
be useful for mechanical processing of metallic glasses and additive
manufacturing.

\vskip 0.5in


\end{abstract}

\maketitle

\section{Introduction}

Understanding the structure evolution in amorphous alloys during
thermal and mechanical treatments is important for tuning their
physical and mechanical properties~\cite{Greer16}.  It is well
accepted by now that in contrast to crystalline solids where
plasticity is governed by topological line defects, known as
disclinations, the elementary plastic events in  amorphous materials
involve collective rearrangements of a few tens of atoms or the
so-called shear transformations~\cite{Spaepen77,Argon79}.   In a
driven system, these rearrangements can assemble into shear bands
where flow becomes sharply localized and act as a precursor for
fracture~\cite{Wang15,Zhong16,Zaccone17,Scudino17}.  Once a shear
band is formed, the structural integrity can be recovered either by
heating a sample above the glass transition temperature and then
cooling back to the glass phase (resetting the structure) or,
alternatively, via mechanical agitation.  For example, it was shown
using atomistic simulations that cracks in nanocrystalline metals
can be completely healed via formation of wedge disclinations during
stress-driven grain boundary migration~\cite{Demkowicz13}. It was
also found experimentally and by means of atomistic simulations that
after steady deformation of bulk metallic glasses, the shear bands
relax during annealing below the glass transition temperature and
the local diffusion coefficient exhibits a nonmonotonic
behavior~\cite{Binkowski16}. In the case of amorphous solids,
however, the effects of periodic loading and initial glass stability
on structural relaxation within the shear band domain, degree of
annealing, and change in mechanical properties yet remain to be
understood.

\vskip 0.05in

During the last decade, molecular dynamics simulations were
particularly valuable in elucidating the atomic mechanisms of
structural relaxation, rejuvenation, and yielding in amorphous
materials under periodic loading
conditions~\cite{Lacks04,Priezjev13,
Sastry13,Reichhardt13,Priezjev14,IdoNature15,Priezjev16,Kawasaki16,
Priezjev16a,Sastry17,Priezjev17,OHern17,Priezjev18,Priezjev18a,
NVP18strload,Sastry18,PriMakrho05,PriMakrho09,Sastry19band,
PriezSHALT19,Ido2020,Priez20ba,Peng20,Jana20,Kawasaki20,KawBer20,BhaSastry20,
Priez20alt,Pelletier20,Priez20del}. Remarkably, it was found that in
athermal, disordered solids subjected to oscillatory shear in the
elastic range, the trajectories of atoms after a number of transient
cycles become exactly reversible and fall into the so-called `limit
cycles'~\cite{Reichhardt13,IdoNature15}.   On the other hand, in the
presence of thermal fluctuations, the relaxation process generally
continues during thousands of cycles and the decay of the potential
energy becomes progressively slower over
time~\cite{Priezjev18,NVP18strload,PriezSHALT19}.   More recently,
it was shown that the critical strain amplitude increases in more
stable athermal glasses~\cite{KawBer20,BhaSastry20}, whereas the
yielding transition can be significantly delayed in mechanically
annealed binary glasses at finite temperature~\cite{Priez20del}.  In
general, the formation of a shear band during the yielding
transition is accelerated in more rapidly annealed glasses
periodically loaded at a higher strain amplitude or when the shear
orientation is alternated in two or three spatial
dimensions~\cite{Priezjev17,Priezjev18a,Sastry19band,
Priez20ba,Priez20alt}.  Interestingly, after a shear band is formed
during cyclic loading, the glass outside the band remains well
annealed, and upon reducing strain amplitude below yield, the
initial shear band anneals out, which leads to reversible dynamics
in the whole domain~\cite{Sastry19band}.   However, despite
extensive efforts, it remains unclear whether mechanical annealing
of a shear band or a crack in metallic glasses depends on the
preparation history, sample size and loading conditions.

\vskip 0.05in

In this paper, the influence of periodic shear deformation in the
elastic range on shear band annealing and mechanical properties of
binary glasses is studied using molecular dynamics simulations. The
system-spanning shear band is initially formed in stable glasses
that were either thermally or mechanically annealed. It will be
shown that small-amplitude oscillatory shear anneals out the shear
band and leads to nearly reversible deformation after a few hundred
cycles at finite temperature. Moreover, upon loading at higher
strain amplitudes, the glasses become increasingly better annealed,
which results in higher yield stress.

\vskip 0.05in

The rest of the paper is outlined as follows. The preparation
procedure, deformation protocol as well as the details of the
simulation model are described in the next section. The time
dependence of the potential energy, mechanical properties, and
spatial organization of atoms with large nonaffine displacements are
presented in section\,\ref{sec:Results}.   The results are briefly
summarized in the last section.

\section{Molecular dynamics (MD) simulations}
\label{sec:MD_Model}

In the present study, the amorphous alloy is represented by the
binary (80:20) Lennard-Jones (LJ) mixture originally introduced by
Kob and Andersen (KA) about twenty years ago~\cite{KobAnd95}. In
this model, the interaction between different types of atoms is
strongly non-additive, thus, allowing formation of a disordered
structure upon slow cooling below the glass transition
temperature~\cite{KobAnd95}. More specifically, the pairwise
interaction is modeled via the LJ potential, as follows:
\begin{equation}
V_{\alpha\beta}(r)=4\,\varepsilon_{\alpha\beta}\,\Big[\Big(\frac{\sigma_{\alpha\beta}}{r}\Big)^{12}\!-
\Big(\frac{\sigma_{\alpha\beta}}{r}\Big)^{6}\,\Big],
\label{Eq:LJ_KA}
\end{equation}
with the parameters: $\varepsilon_{AA}=1.0$, $\varepsilon_{AB}=1.5$,
$\varepsilon_{BB}=0.5$, $\sigma_{AA}=1.0$, $\sigma_{AB}=0.8$,
$\sigma_{BB}=0.88$, and $m_{A}=m_{B}$~\cite{KobAnd95}. It should be
mentioned that a similar parametrization was used by Weber and
Stillinger to study the amorphous metal-metalloid alloy
$\text{Ni}_{80}\text{P}_{20}$~\cite{Weber85}. To save computational
time, the LJ potential was truncated at the cutoff radius
$r_{c,\,\alpha\beta}=2.5\,\sigma_{\alpha\beta}$. The total number of
atoms is fixed $N=60\,000$ throughout the study.   For clarity, all
physical quantities are reported in terms of the reduced units of
length, mass, and energy $\sigma=\sigma_{AA}$, $m=m_{A}$, and
$\varepsilon=\varepsilon_{AA}$.  Using the LAMMPS parallel code, the
equations of motion were integrated via the velocity Verlet
algorithm with the time step $\triangle t_{MD}=0.005\,\tau$, where
$\tau=\sigma\sqrt{m/\varepsilon}$ is the LJ
time~\cite{Allen87,Lammps}.

\vskip 0.05in


All simulations were carried out at a constant density
$\rho=\rho_A+\rho_B=1.2\,\sigma^{-3}$ in a periodic box of linear
size $L=36.84\,\sigma$. It was previously found that the computer
glass transition temperature of the KA model at the density
$\rho=1.2\,\sigma^{-3}$ is
$T_c=0.435\,\varepsilon/k_B$~\cite{KobAnd95}. The system temperature
was maintained via the Nos\'{e}-Hoover
thermostat~\cite{Allen87,Lammps}.  After thorough equilibration and
gradual annealing at the temperature $T_{LJ}=0.01\,\varepsilon/k_B$,
the system was subjected to periodic shear deformation along the
$xz$ plane as follows:
\begin{equation}
\gamma(t)=\gamma_0\,\text{sin}(2\pi t/T),
\label{Eq:shear}
\end{equation}
where $\gamma_0$ is the strain amplitude and $T=5000\,\tau$ is the
period of oscillation. The corresponding oscillation frequency is
$\omega=2\pi/T=1.26\times10^{-3}\,\tau^{-1}$. Once a shear band was
formed at $\gamma_0=0.080$, the glasses were periodically strained
at the strain amplitudes $\gamma_0=0.030$, $0.040$, $0.050$,
$0.060$, and $0.065$ during 3000 cycles. It was previously found
that in the case of poorly annealed (rapidly cooled) glasses, the
critical value of the strain amplitude at the temperature
$T_{LJ}=0.01\,\varepsilon/k_B$ and density $\rho=1.2\,\sigma^{-3}$
is $\gamma_0\approx0.067$~\cite{Priez20alt}. The typical simulation
during 3000 cycles takes about 80 days using 40 processors in
parallel.

\vskip 0.05in

\begin{figure}[t]
\includegraphics[width=9.0cm,angle=0]{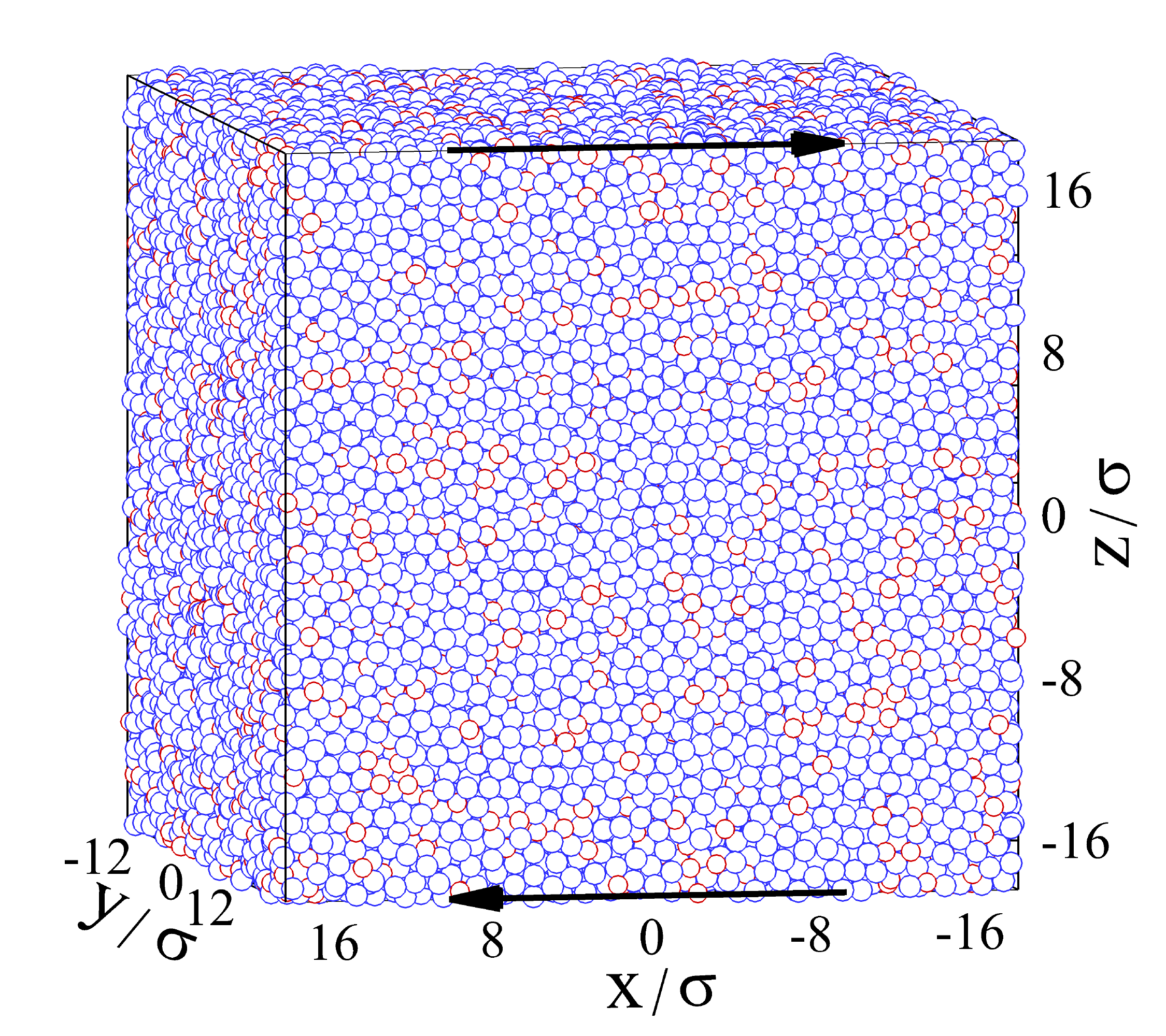}
\caption{(Color online) A snapshot of the \textit{thermally
annealed} glass at the temperature $T_{LJ}=0.01\,\varepsilon/k_B$.
The system consists of 48\,000 atoms of type \textit{A} (large blue
circles) and 12\,000 atoms of type \textit{B} (small red circles) in
a periodic box of linear size $L=36.84\,\sigma$. Atoms are not shown
to scale. The black arrows indicate the direction of oscillatory
shear deformation along the $xz$ plane. }
\label{fig:shapshot}
\end{figure}

For the simulation results presented in the next section, the
preparation and the initial loading protocols are the same as the
ones used in the previous MD study on the yielding transition in
stable glasses~\cite{Priez20del}. Briefly, the binary mixture was
first equilibrated at $T_{LJ}=1.0\,\varepsilon/k_B$ and
$\rho=1.2\,\sigma^{-3}$ and then slowly cooled with the rate
$10^{-5}\varepsilon/k_{B}\tau$ to $T_{LJ}=0.30\,\varepsilon/k_B$.
Furthermore, one sample was cooled down to
$T_{LJ}=0.01\,\varepsilon/k_B$ during the time interval $10^4\,\tau$
(see Fig.\,\ref{fig:shapshot}).  The other sample was mechanically
annealed at $T_{LJ}=0.30\,\varepsilon/k_B$ via cyclic loading at
$\gamma_0=0.035$ during 600 cycles, and only then cooled to
$T_{LJ}=0.01\,\varepsilon/k_B$ during $10^4\,\tau$.   Thus, after
relocating glasses to $T_{LJ}=0.01\,\varepsilon/k_B$, two glass
samples with different processing history and potential energies
were obtained. In what follows, these samples will be referred to as
\textit{thermally annealed} and \textit{mechanically annealed}
glasses.

\section{Results}
\label{sec:Results}


Amorphous alloys typically undergo physical aging, when a system
slowly evolves towards lower energy states, and generally this
process can be accelerated by external cyclic deformation within the
elastic range~\cite{Qiao19}. Thus, the structural relaxation of
disordered solids under periodic loading proceeds via collective,
irreversible rearrangements of
atoms~\cite{Sastry13,Priezjev18,Priezjev18a,Sastry18}, while at
sufficiently low energy levels, mechanical annealing becomes
inefficient~\cite{KawBer20}. The two glass samples considered in the
present study were prepared either via mechanical annealing at a
temperature not far below the glass transition temperature or by
computationally slow cooling from the liquid state. It was
previously shown that small-amplitude periodic shear deformation at
temperatures well below $T_g$ does not lead to further annealing of
these glasses~\cite{Priez20del}.  Rather, the results presented
below focus on the annealing process of a shear band, introduced in
these samples by large periodic strain, and subsequent recovery of
their mechanical properties.

\vskip 0.05in

\begin{figure}[t]
\includegraphics[width=12.0cm,angle=0]{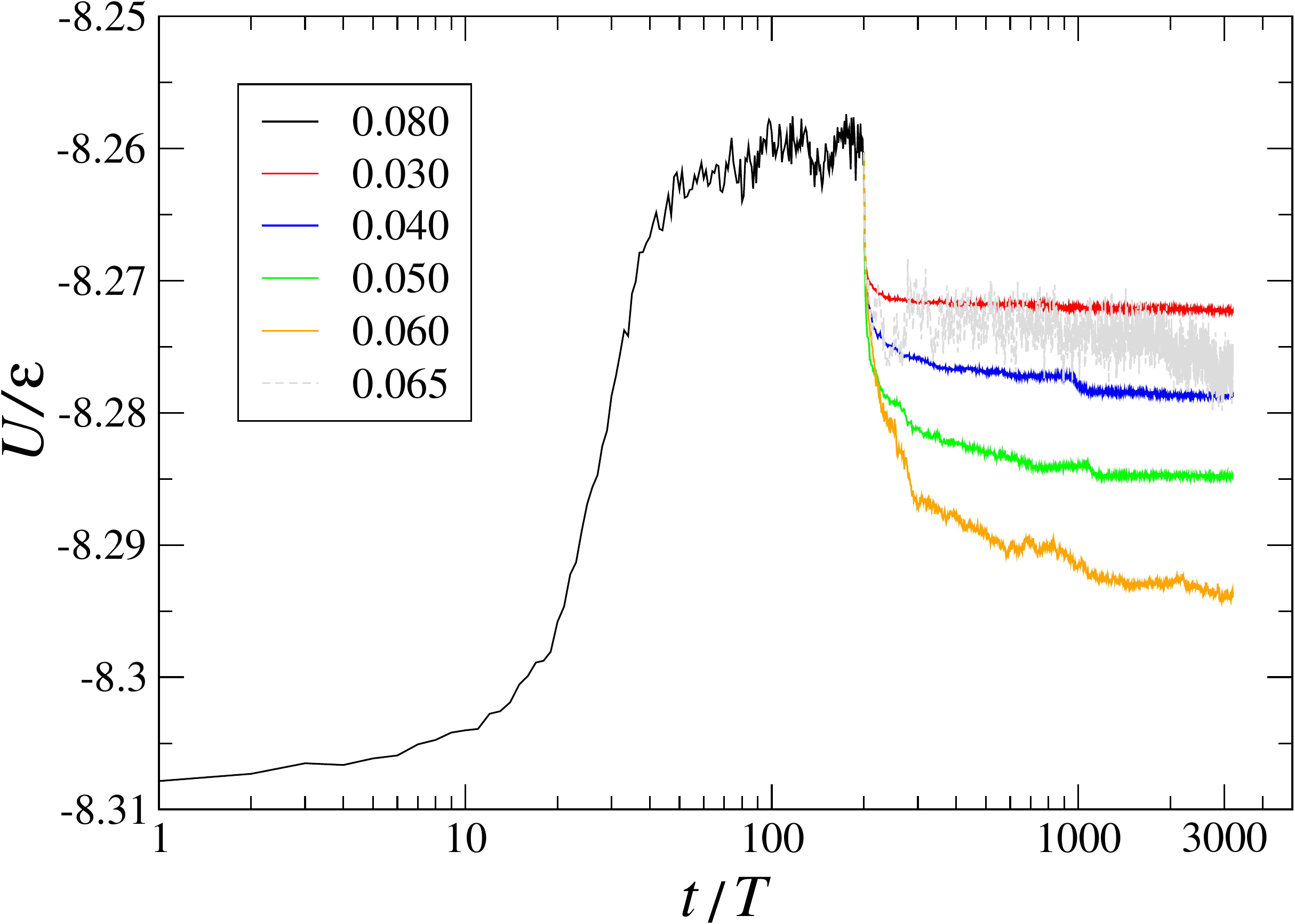}
\caption{(Color online) The dependence of the potential energy
minima (at zero strain) on the number of cycles for the indicated
values of the strain amplitude.  The shear band was formed in the
\textit{thermally annealed} glass during the first 200 cycles at the
strain amplitude $\gamma_0=0.080$ (the black curve). The system
temperature is $T_{LJ}=0.01\,\varepsilon/k_B$ and the oscillation
period is $T=5000\,\tau$. }
\label{fig:poten_Quench_SB_heal}
\end{figure}

The time dependence of the potential energy at the end of each cycle
is reported in Fig.\,\ref{fig:poten_Quench_SB_heal} for the
\textit{thermally annealed} glass. In this case, the glass was first
subjected to oscillatory shear during 200 cycles with the strain
amplitude $\gamma_0=0.080$ (see the black curve in
Fig.\,\ref{fig:poten_Quench_SB_heal}). The strain amplitude
$\gamma_0=0.080$ is slightly larger than the critical strain
amplitude $\gamma_0\approx0.067$ at $T_{LJ}=1.0\,\varepsilon/k_B$
and $\rho=1.2\,\sigma^{-3}$~\cite{Priez20alt}, and, therefore, the
periodic loading induced the formation of a shear band across the
system after about 20 cycles. As shown in
Fig.\,\ref{fig:poten_Quench_SB_heal}, the process of shear band
formation is associated with a sharp increase in the potential
energy followed by a plateau at $U\approx-8.26\,\varepsilon$ with
pronounced fluctuations due to plastic flow within the band.  It was
previously demonstrated that during the plateau period, the periodic
deformation involves two well separated domains with diffusive and
reversible dynamics~\cite{Priez20del}.

\vskip 0.05in


After the shear band became fully developed in the \textit{thermally
annealed} glass, the strain amplitude of periodic deformation was
reduced in the range $0.030\leqslant \gamma_0 \leqslant 0.065$ when
$t=200\,T$. The results in Fig.\,\ref{fig:poten_Quench_SB_heal}
indicate that the potential energy of the system is gradually
reduced when $t>200\,T$, and the energy drop increases at higher
strain amplitudes (except for $\gamma_0=0.065$). Notice that the
potential energy levels out at $t\gtrsim 1300\,T$ for
$\gamma_0=0.030$, $0.040$, and $0.050$, while the relaxation process
continues up to $t=3200\,T$ for $\gamma_0=0.060$.  These results
imply that the shear band becomes effectively annealed by the
small-amplitude oscillatory shear, leading to nearly reversible
dynamics in the whole sample, as will be illustrated below via the
analysis of nonaffine displacements. By contrast, the deformation
within the shear band remains irreversible at the higher strain
amplitude $\gamma_0=0.065$ (denoted by the fluctuating grey curve in
Fig.\,\ref{fig:poten_Quench_SB_heal}). This observation can be
rationalized by realizing that the strain remains localized within
the shear band, and the effective strain amplitude within the band
is greater than the critical value
$\gamma_0\approx0.067$~\cite{Priez20alt}.

\vskip 0.05in

\begin{figure}[t]
\includegraphics[width=12.0cm,angle=0]{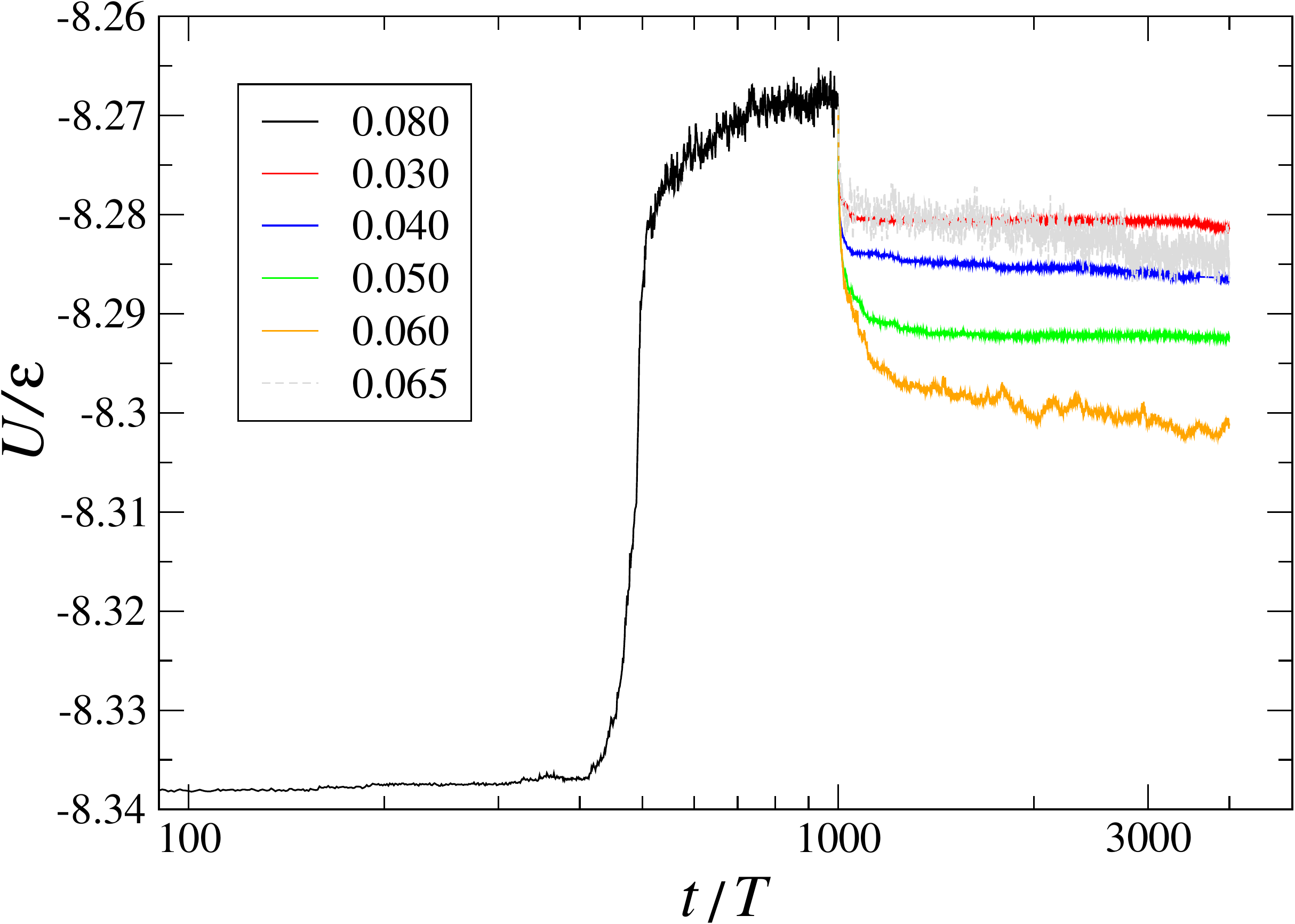}
\caption{(Color online) The variation of the potential energy (at
the end of each cycle) as a function of the cycle number for the
selected strain amplitudes. The shear band was introduced in the
\textit{mechanically annealed} glass after 1000 cycles at the strain
amplitude $\gamma_0=0.080$ (the black curve; see text for details).
The time is reported in terms of oscillation periods, \textit{i.e.},
$T=5000\,\tau$. The temperature is $T_{LJ}=0.01\,\varepsilon/k_B$. }
\label{fig:poten_600cyc_SB_heal}
\end{figure}

The potential energy minima for the \textit{mechanically annealed}
glass are presented in Fig.\,\ref{fig:poten_600cyc_SB_heal} for the
indicated strain amplitudes.  It should be commented that the
preparation protocol, which included 600 cycles at $\gamma_0=0.035$
and $T_{LJ}=0.30\,\varepsilon/k_B$, produced an atomic configuration
with a relatively deep potential energy level, \textit{i.e.},
$U\approx-8.337\,\varepsilon$. Upon periodic loading at
$\gamma_0=0.080$ and $T_{LJ}=0.01\,\varepsilon/k_B$, the yielding
transition is delayed by about 450 cycles, as shown by the black
curve in Fig.\,\ref{fig:poten_600cyc_SB_heal} (the same data as in
Ref.\,\cite{Priez20del}).   Similarly to the case of thermally
annealed glasses, the potential energy in
Fig.\,\ref{fig:poten_600cyc_SB_heal} is gradually reduced when the
strain amplitude is changed from $\gamma_0=0.080$ to the selected
values in the range $0.030\leqslant \gamma_0 \leqslant 0.065$.
Interestingly, the largest decrease in the potential energy at the
strain amplitude $\gamma_0=0.060$ is nearly the same ($\Delta
U\approx 0.03\,\varepsilon$) for both thermally and mechanically
annealed glasses. In addition, it can be commented that in both
cases presented in Figs.\,\ref{fig:poten_Quench_SB_heal} and
\ref{fig:poten_600cyc_SB_heal}, the potential energy remains above
the energy levels of initially stable glasses (before a shear band
is formed) even for loading at the strain amplitude
$\gamma_0=0.060$. The results of a previous MD study on mechanical
annealing of \textit{rapidly quenched} glasses imply that the energy
level $U\approx-8.31\,\varepsilon$ can be reached via cyclic loading
at $T_{LJ}=0.01\,\varepsilon/k_B$ but it might take thousands of
additional cycles~\cite{Priez20alt}.


\vskip 0.05in


While the potential energy within a shear band becomes relatively
large, the energy of the glass outside the band remains largely
unaffected during the yielding transition.  As shown above, the
\textit{mechanically annealed} glass is initially more stable (has a
lower potential energy) than the \textit{thermally annealed} glass.
This in turn implies that the boundary conditions for the subyield
loading of the shear band are different in the two cases, and,
therefore, the potential energy change during the relaxation
process, in principle, might also vary.  In other words, the
annealing of the shear band by small-amplitude periodic deformation
might be affected by the atomic structure of the adjacent glass.
However, the results in Figs.\,\ref{fig:poten_Quench_SB_heal} and
\ref{fig:poten_600cyc_SB_heal} suggest that the potential energy
change is roughly the same in both cases; although a more careful
analysis might be needed in the future to clarify this point.

\vskip 0.05in

\begin{figure}[t]
\includegraphics[width=12.0cm,angle=0]{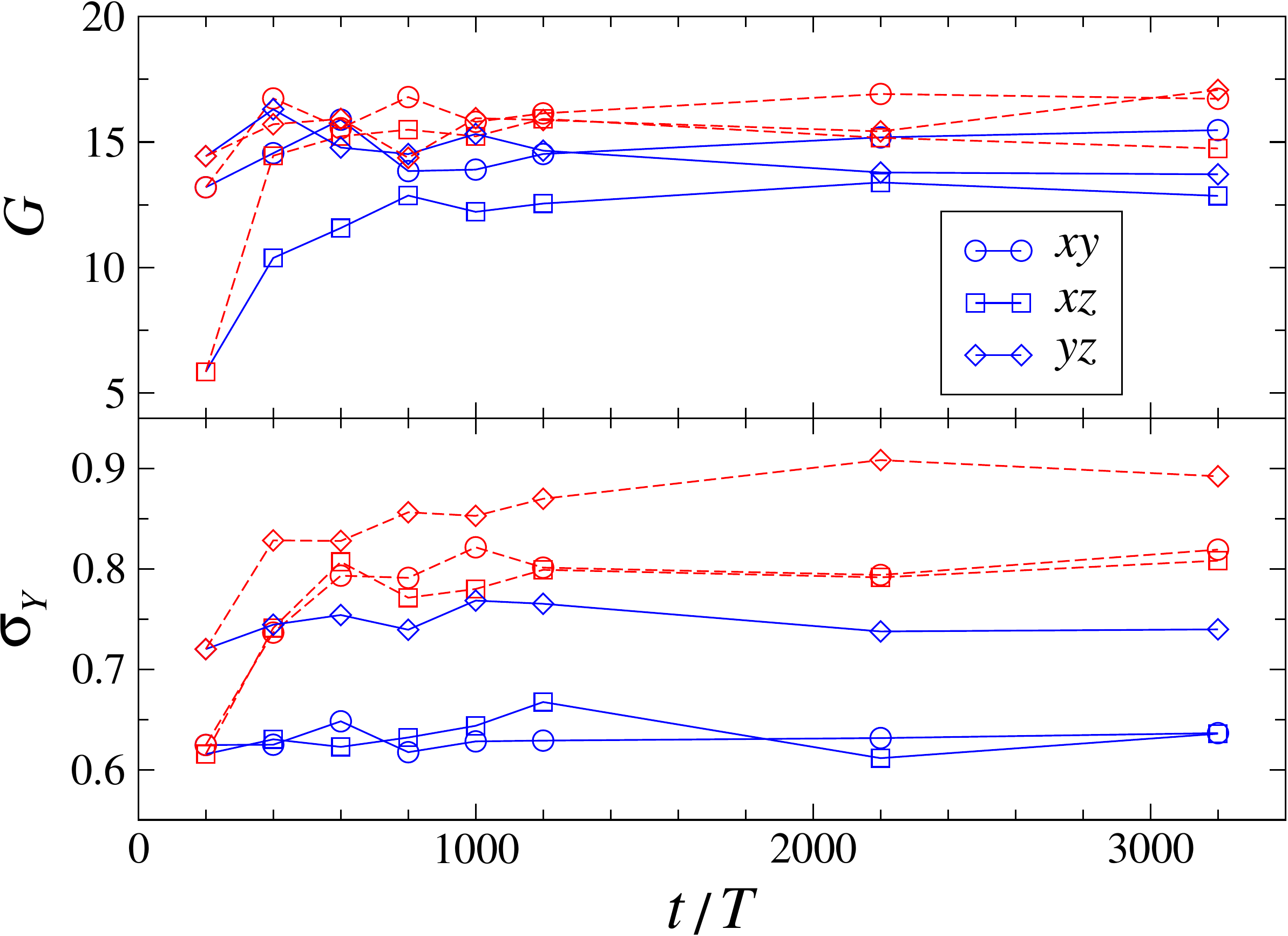}
\caption{(Color online)  The shear modulus $G$ (in units of
$\varepsilon\sigma^{-3}$) and yielding peak $\sigma_Y$ (in units of
$\varepsilon\sigma^{-3}$) as a function of the cycle number for the
\textit{thermally annealed} glass.  The startup continuous shear
with the strain rate $\dot{\gamma}=10^{-5}\,\tau^{-1}$ was applied
along the $xy$ plane (circles), $xz$ plane (squares), and $yz$ plane
(diamonds).  Before startup deformation, the samples were
periodically deformed with the strain amplitudes $\gamma_0=0.030$
(solid blue) and $\gamma_0=0.060$ (dashed red). The time range is
the same as in Fig.\,\ref{fig:poten_Quench_SB_heal}. }
\label{fig:G_and_Y_thermq}
\end{figure}
\begin{figure}[t]
\includegraphics[width=12.0cm,angle=0]{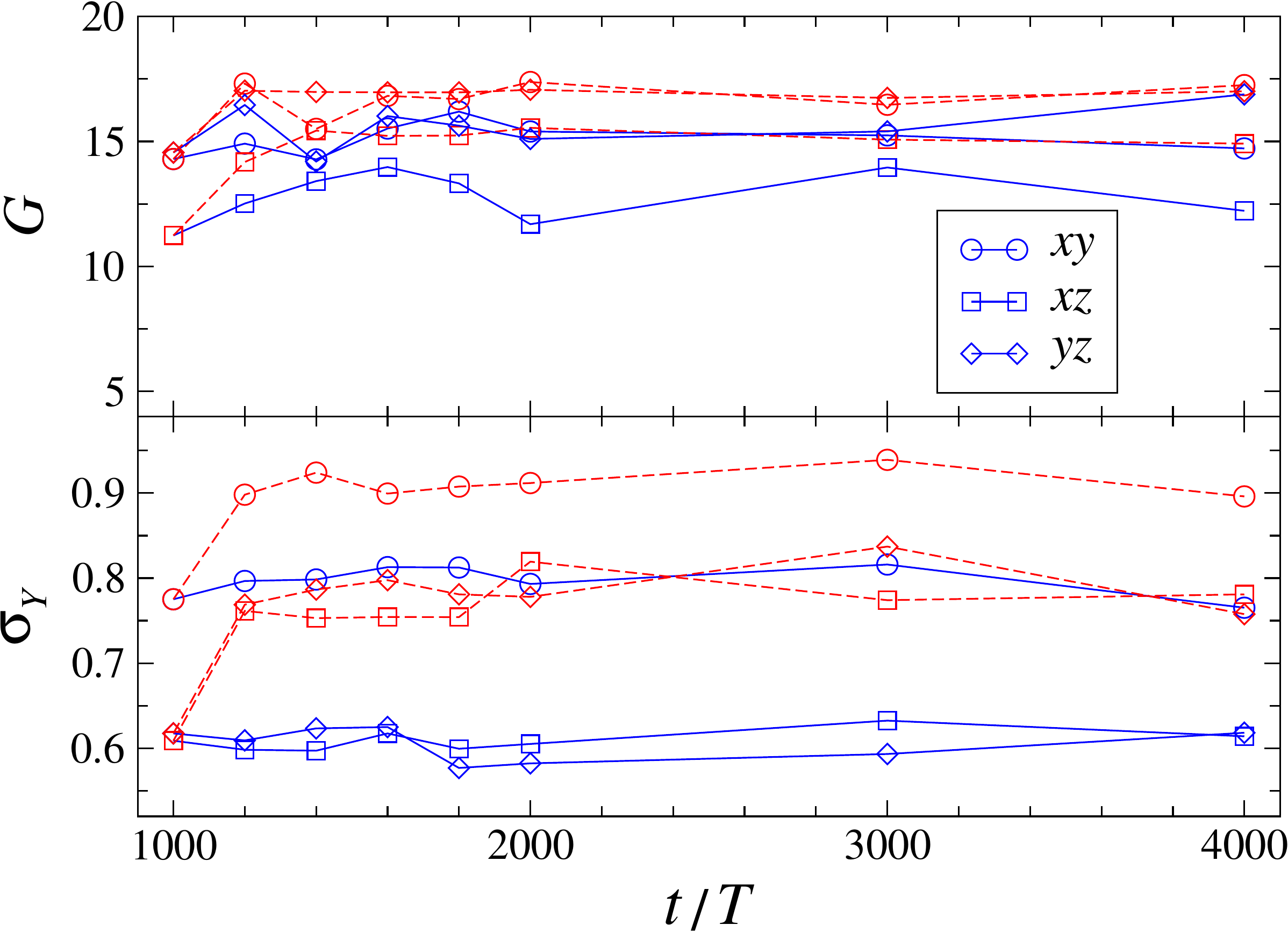}
\caption{(Color online)  The shear modulus $G$ (in units of
$\varepsilon\sigma^{-3}$) and yielding peak $\sigma_Y$ (in units of
$\varepsilon\sigma^{-3}$) versus cycle number for the
\textit{mechanically annealed} glass. The startup shear deformation
with the strain rate $\dot{\gamma}=10^{-5}\,\tau^{-1}$ was imposed
along the $xy$ plane (circles), $xz$ plane (squares), and $yz$ plane
(diamonds). Before continuous shear, the samples were cyclically
deformed with the strain amplitudes $\gamma_0=0.030$ (solid blue)
and $\gamma_0=0.060$ (dashed red). The same cycle range as in
Fig.\,\ref{fig:poten_600cyc_SB_heal}. }
\label{fig:G_and_Y_600cyc}
\end{figure}

We next report the results of mechanical tests that involve startup
continuous shear deformation in order to probe the effect of
small-amplitude periodic loading on the yield stress.   The shear
modulus, $G$, and the peak value of the stress overshoot,
$\sigma_Y$, are plotted in Figs.\,\ref{fig:G_and_Y_thermq} and
\ref{fig:G_and_Y_600cyc} for glasses that were periodically deformed
with the strain amplitudes $\gamma_0=0.030$ and $0.060$.   In each
case, the startup deformation was imposed along the $xy$, $xz$, and
$yz$ planes with the constant strain rate
$\dot{\gamma}=10^{-5}\,\tau^{-1}$. The data are somewhat scattered,
since simulations were carried out only for one realization of
disorder, but the trends are evident. First, both $G$ and $\sigma_Y$
are relatively small when shear is applied along the $xz$ plane at
$t=200\,T$ in Fig.\,\ref{fig:G_and_Y_thermq} and at $t=1000\,T$ in
Fig.\,\ref{fig:G_and_Y_600cyc} because of the shear band that was
formed previously at $\gamma_0=0.080$.  Second, the shear modulus
and yield stress increase towards plateau levels during the next few
hundred cycles, and their magnitudes are greater for the larger
strain amplitude $\gamma_0=0.060$, since those samples were annealed
to deeper energy states (see Figs.\,\ref{fig:poten_Quench_SB_heal}
and \ref{fig:poten_600cyc_SB_heal}).

\vskip 0.05in


The results in Figures\,\ref{fig:G_and_Y_thermq}\,(b) and
\ref{fig:G_and_Y_600cyc}\,(b) show that the yield stress is only
weakly dependent on the number of cycles in glasses that were
periodically strained at the smaller amplitude $\gamma_0=0.030$,
whereas for $\gamma_0=0.060$, the yield stress increases noticeably
and levels out at $\sigma_Y\approx0.9\,\varepsilon\sigma^{-3}$ for
the \textit{mechanically annealed} glass and at a slightly smaller
value for the \textit{thermally annealed} glass. It was previously
shown that the yield stress is slightly larger, \textit{i.e.},
$\sigma_Y\approx1.05\,\varepsilon\sigma^{-3}$, for rapidly quenched
glasses that were mechanically annealed at the strain amplitude
$\gamma_0=0.060$ for similar loading conditions~\cite{PriezSHALT19}.
This discrepancy might arise because in Ref.\,\cite{PriezSHALT19}
the glass was homogenously annealed starting from the rapidly
quenched state, while in the present study, the potential energy
within the annealed shear-band domain always remains higher than in
the rest of the sample, thus resulting in spatially heterogeneous
structure.   On the other hand, it was recently shown that the
presence of an interface between relaxed and rejuvenated domains in
a relatively large sample might impede strain
localization~\cite{Kosiba19}.

\vskip 0.05in


The relative rearrangements of atoms with respect to their neighbors
in a deformed amorphous system can be conveniently quantified via
the so-called nonaffine displacements.  By definition, the nonaffine
measure $D^2(t, \Delta t)$ for an atom $i$ is computed via the
transformation matrix $\mathbf{J}_i$ that minimizes the following
expression for a group of neighboring atoms:
\begin{equation}
D^2(t, \Delta t)=\frac{1}{N_i}\sum_{j=1}^{N_i}\Big\{
\mathbf{r}_{j}(t+\Delta t)-\mathbf{r}_{i}(t+\Delta t)-\mathbf{J}_i
\big[ \mathbf{r}_{j}(t) - \mathbf{r}_{i}(t)  \big] \Big\}^2,
\label{Eq:D2min}
\end{equation}
where $\Delta t$ is the time interval between two atomic
configurations, and the summation is performed over the nearest
neighbors located within $1.5\,\sigma$ from the position of the
$i$-th atom at $\mathbf{r}_{i}(t)$. The nonaffine quantity defined
by Eq.\,(\ref{Eq:D2min}) was originally introduced by Falk and
Langer in order to accurately detect the localized shear
transformations that involved swift rearrangements of small groups
of atoms in driven disordered solids~\cite{Falk98}. In the last few
years, this method was widely used to study the collective,
irreversible dynamics of atoms in binary glasses subjected to time
periodic~\cite{Priezjev16,Priezjev16a,Priezjev17,Priezjev18,Priezjev18a,
PriezSHALT19,Priez20ba,Peng20,KawBer20,Priez20alt} and startup
continuous~\cite{HorbachJR16,Schall07,Pastewka19,Priez20tfic,Priez19star,Ozawa20,ShiBai20}
shear deformation, tension-compression cyclic
loading~\cite{NVP18strload,Jana20}, prolonged elastostatic
compression~\cite{PriezELAST19,PriezELAST20}, creep~\cite{Eckert21}
and thermal cyclic
loading~\cite{Priez19one,Priez19tcyc,Priez19T2000,Priez19T5000,Guan20}.

\vskip 0.05in

\begin{figure}[t]
\includegraphics[width=12.0cm,angle=0]{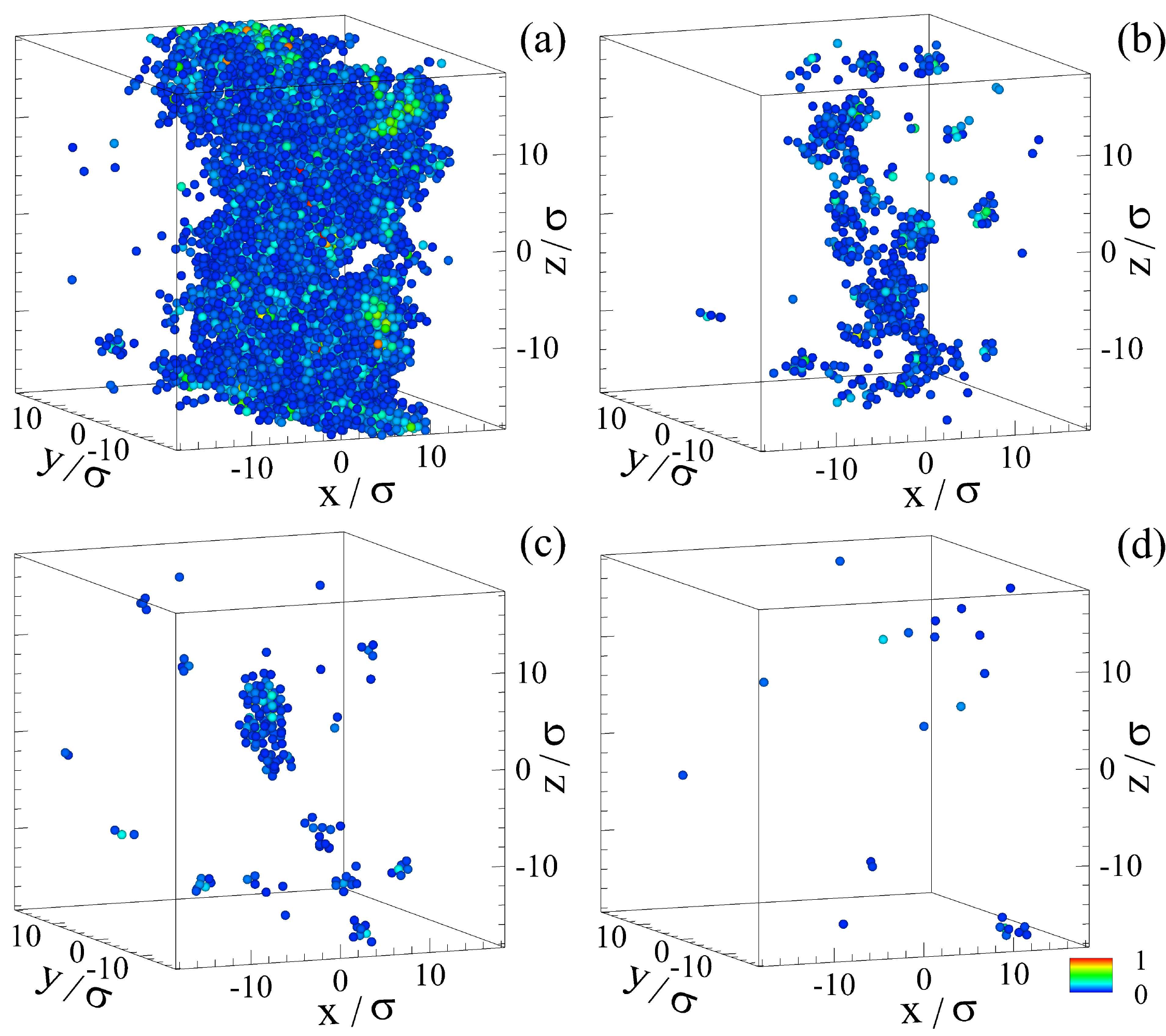}
\caption{(Color online) A series of snapshots of atomic
configurations during periodic shear with the strain amplitude
$\gamma_0=0.030$. The loading conditions are the same as in
Fig.\,\ref{fig:poten_Quench_SB_heal} (the red curve). The nonaffine
measure in Eq.\,(\ref{Eq:D2min}) is (a) $D^2(200\,T,
T)>0.04\,\sigma^2$, (b) $D^2(205\,T, T)>0.04\,\sigma^2$, (c)
$D^2(220\,T, T)>0.04\,\sigma^2$, and (d) $D^2(300\,T,
T)>0.04\,\sigma^2$. The colorcode in the legend denotes the
magnitude of $D^2$. Atoms are not shown to scale.  }
\label{fig:snapshots_Tquench_T001_amp080_heal_amp030_1_5_20_100}
\end{figure}
\begin{figure}[t]
\includegraphics[width=12.0cm,angle=0]{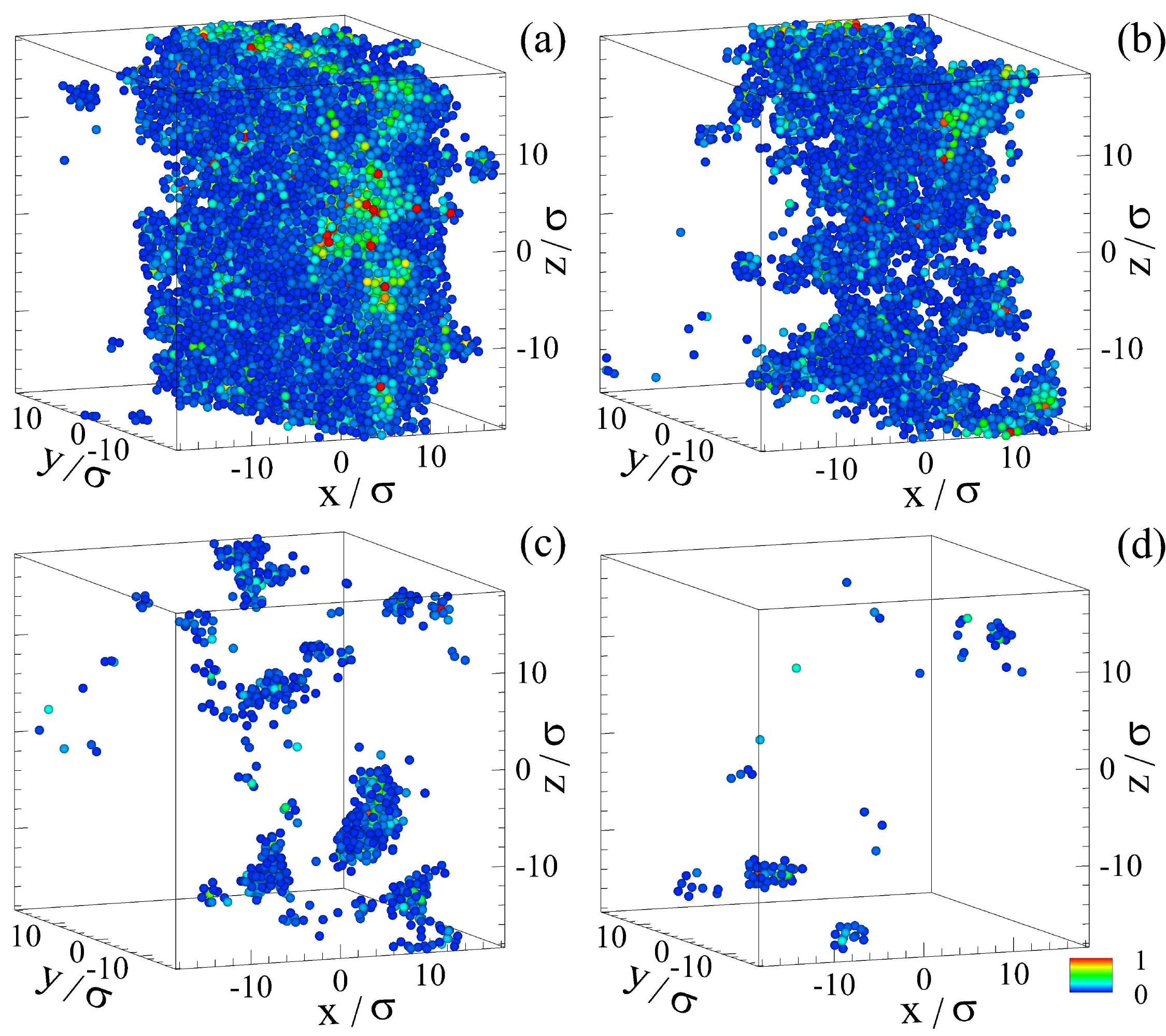}
\caption{(Color online) The position of atoms in the thermally
annealed glass subjected to periodic shear with the strain amplitude
$\gamma_0=0.060$. The corresponding potential energy is denoted by
the orange curve in Fig.\,\ref{fig:poten_Quench_SB_heal}.  The
nonaffine measure is (a) $D^2(200\,T, T)>0.04\,\sigma^2$, (b)
$D^2(220\,T, T)>0.04\,\sigma^2$, (c) $D^2(300\,T,
T)>0.04\,\sigma^2$, and (d) $D^2(1200\,T, T)>0.04\,\sigma^2$. The
magnitude of $D^2$ is defined in the legend.   }
\label{fig:snapshots_Tquench_T001_amp080_heal_amp060_1_20_100_1000}
\end{figure}

The representative snapshots of \textit{thermally annealed} glasses
are presented in
Fig.\,\ref{fig:snapshots_Tquench_T001_amp080_heal_amp030_1_5_20_100}
for the strain amplitude $\gamma_0=0.030$ and in
Fig.\,\ref{fig:snapshots_Tquench_T001_amp080_heal_amp060_1_20_100_1000}
for $\gamma_0=0.060$. For clarity, only atoms with relatively large
nonaffine displacements during one oscillation period are displayed.
Note that the typical cage size at $\rho=1.2\,\sigma^{-3}$ is about
$0.1\,\sigma$~\cite{Priezjev13}, and, therefore, the displacements
of atoms with $D^2(n\,T, T)>0.04\,\sigma^2$ correspond to
cage-breaking events. It can be clearly seen in the panel (a) of
Figures\,\ref{fig:snapshots_Tquench_T001_amp080_heal_amp030_1_5_20_100}
and
\ref{fig:snapshots_Tquench_T001_amp080_heal_amp060_1_20_100_1000},
that the shear band runs along the $yz$ plane right after switching
to the subyield loading regime.  As expected, the magnitude of
$D^2(200\,T, T)$ on average decays towards the interfaces. Upon
continued loading, the shear band becomes thinner and eventually
breaks up into isolated clusters whose size is reduced over time.
The coarsening process is significantly slower for the strain
amplitude $\gamma_0=0.060$ (about 1000 cycles) than for
$\gamma_0=0.030$ (about 200 cycles). This trend is consistent with
the decay of the potential energy denoted in
Fig.\,\ref{fig:poten_Quench_SB_heal} by the red and orange curves.

\vskip 0.05in

\begin{figure}[t]
\includegraphics[width=12.0cm,angle=0]{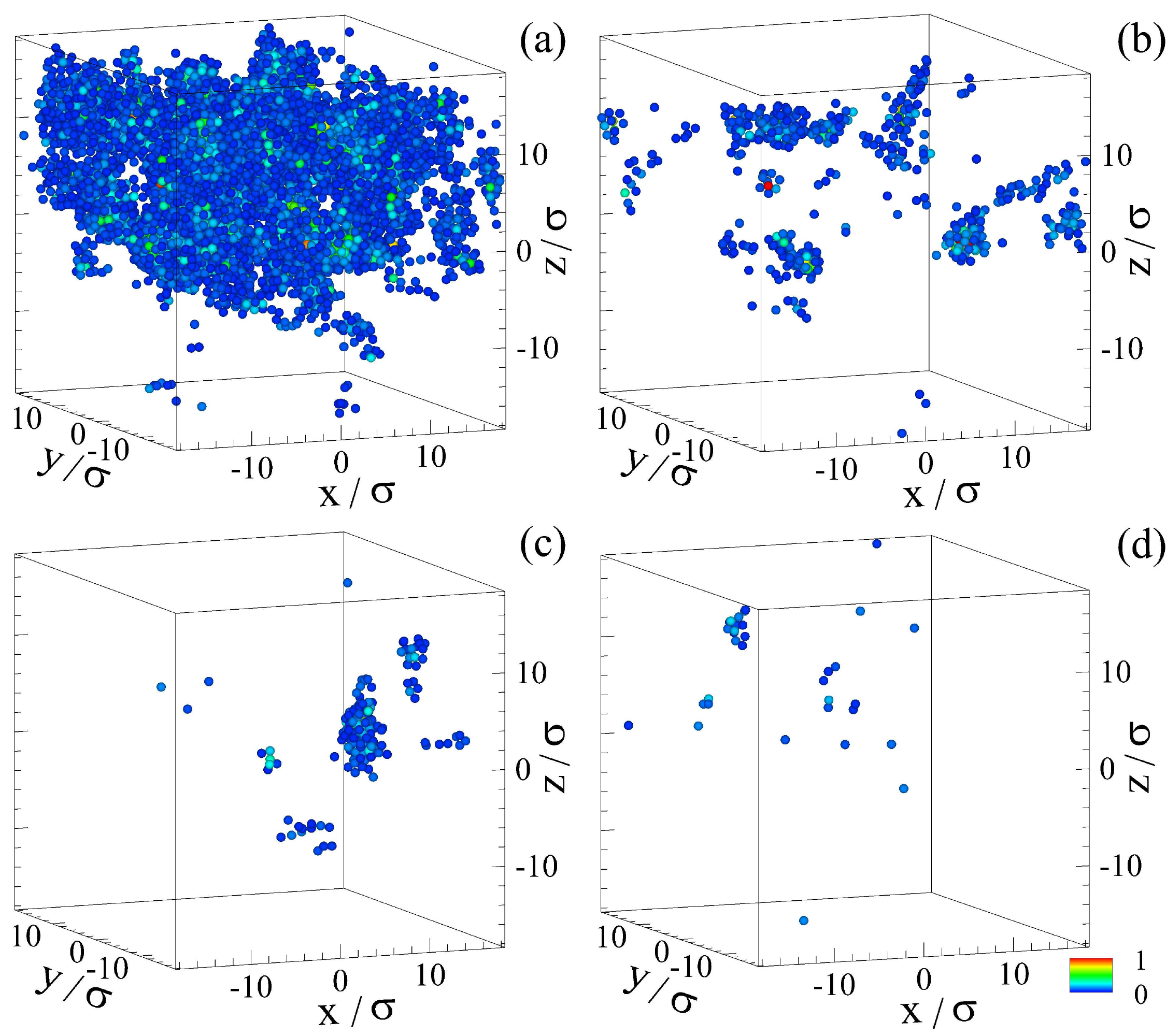}
\caption{(Color online) Instantaneous snapshots of the binary glass
periodically sheared with the strain amplitude $\gamma_0=0.030$. The
data correspond to the red curve in
Fig.\,\ref{fig:poten_600cyc_SB_heal}.  The nonaffine quantity is (a)
$D^2(1000\,T, T)>0.04\,\sigma^2$, (b) $D^2(1005\,T,
T)>0.04\,\sigma^2$, (c) $D^2(1010\,T, T)>0.04\,\sigma^2$, and (d)
$D^2(1100\,T, T)>0.04\,\sigma^2$. The colorcode denotes the
magnitude of $D^2$. }
\label{fig:snapshots_600cyc_T001_amp080_heal_amp030_1_5_10_100}
\end{figure}
\begin{figure}[t]
\includegraphics[width=12.0cm,angle=0]{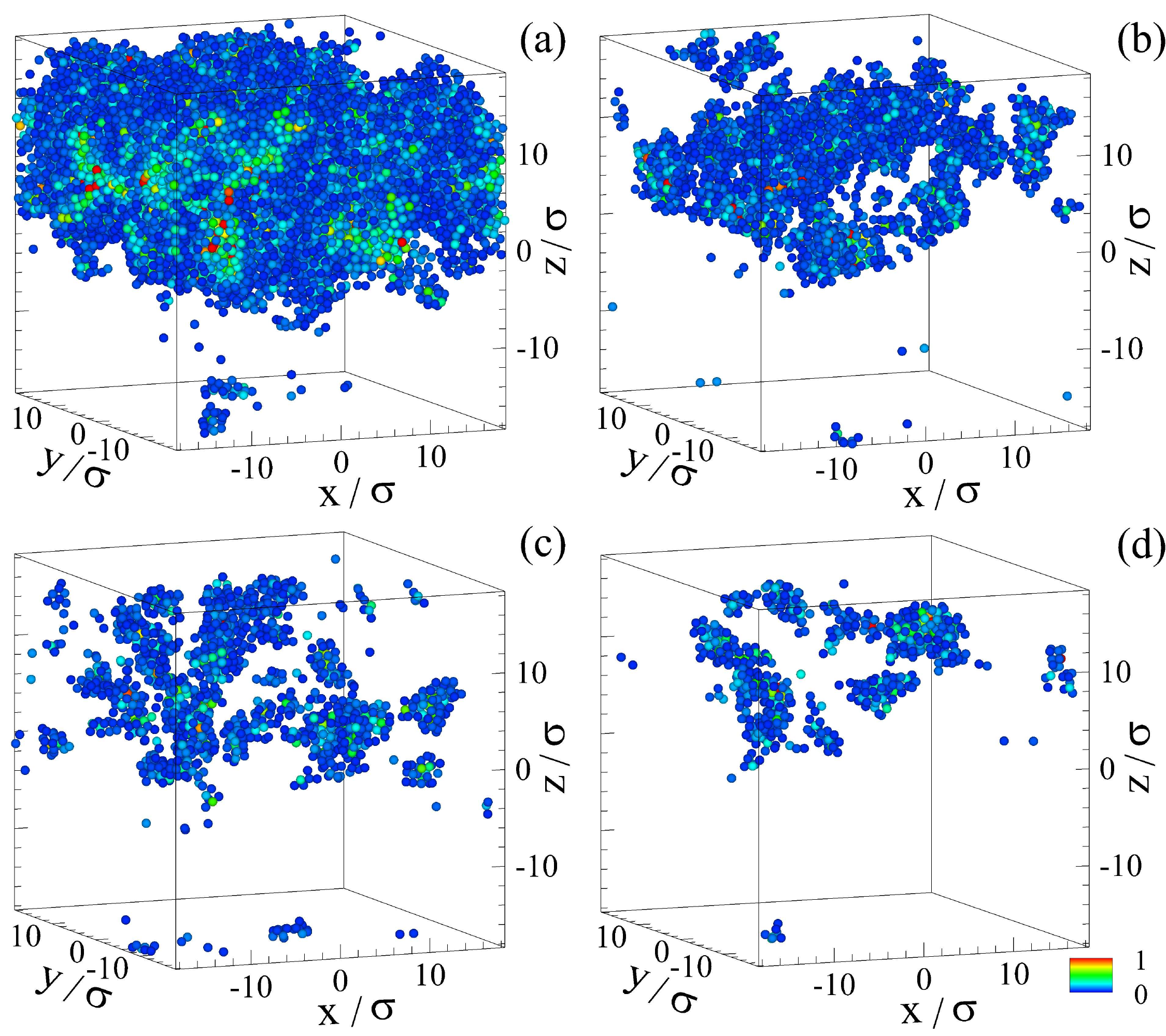}
\caption{(Color online) Atomic positions in the binary glass
cyclically loaded at the strain amplitude $\gamma_0=0.060$. The data
are taken from the selected time intervals along the orange curve in
Fig.\,\ref{fig:poten_600cyc_SB_heal}. The nonaffine quantity is (a)
$D^2(1000\,T, T)>0.04\,\sigma^2$, (b) $D^2(1100\,T,
T)>0.04\,\sigma^2$, (c) $D^2(1200\,T, T)>0.04\,\sigma^2$, and (d)
$D^2(3000\,T, T)>0.04\,\sigma^2$. $D^2$ is defined in the legend. }
\label{fig:snapshots_600cyc_T001_amp080_heal_amp060_1_100_200_2000}
\end{figure}

Similar conclusions can be drawn by visual inspection of consecutive
snapshots of the \textit{mechanically annealed} glass cyclically
loaded at the strain amplitude $\gamma_0=0.030$ (see
Fig.\,\ref{fig:snapshots_600cyc_T001_amp080_heal_amp030_1_5_10_100})
and at $\gamma_0=0.060$ (see
Fig.\,\ref{fig:snapshots_600cyc_T001_amp080_heal_amp060_1_100_200_2000}).
It can be observed that the shear band is initially oriented along
the $xy$ plane, which is consistent with a relatively large value of
the yield stress along the $xy$ direction at $t=1000\,T$ in
Fig.\,\ref{fig:G_and_Y_600cyc}. The atomic trajectories become
nearly reversible already after about 10 cycles at the strain
amplitude $\gamma_0=0.030$, as shown in
Fig.\,\ref{fig:snapshots_600cyc_T001_amp080_heal_amp030_1_5_10_100},
while isolated clusters of atoms with large nonaffine displacements
are still present after about 2000 cycles at $\gamma_0=0.060$ (see
Fig.\,\ref{fig:snapshots_600cyc_T001_amp080_heal_amp060_1_100_200_2000}).
Altogether these results indicate that oscillatory shear deformation
with a strain amplitude just below the critical value can be used to
effectively anneal a shear band and make the amorphous material
stronger.

\section{Conclusions}

In summary, the process of shear band annealing in metallic glasses
subjected to small-amplitude periodic shear deformation was examined
using molecular dynamics simulations. The glass was modeled as a
binary mixture with non-additive interaction between atoms of
different types, and the shear band was initially developed in
stable glasses under oscillatory shear above the yielding point. It
was shown that periodic loading in the elastic range results in a
gradual decay of the potential energy over consecutive cycles, and
upon increasing strain amplitude, lower energy states can be
accessed after thousands of cycles. Furthermore, the spatiotemporal
analysis of nonaffine displacements demonstrated that a shear band
becomes thinner and breaks into separate clusters whose size is
reduced upon continued loading. Thus, in a wide range of strain
amplitudes below yield, the cyclic loading leads to a nearly
reversible dynamics of atoms at finite temperature. Lastly, both the
shear modulus and yield stress saturate to higher values as the
shear band region becomes better annealed at higher strain
amplitudes.

\section*{Acknowledgments}

Financial support from the National Science Foundation (CNS-1531923)
is gratefully acknowledged.  The article was prepared within the
framework of the HSE University Basic Research Program and funded in
part by the Russian Academic Excellence Project `5-100'. The
simulations were performed at Wright State University's Computing
Facility and the Ohio Supercomputer Center. The molecular dynamics
simulations were carried out using the parallel LAMMPS code
developed at Sandia National Laboratories~\cite{Lammps}.



%

%

%

%

%

%

%

%

%

\bibliographystyle{prsty}

\end{document}